\documentclass[prl,twocolumn,showpacs,preprintnumbers,amsmath,amssymb, superscriptaddress]{revtex4}

\usepackage{graphicx}
\usepackage{dcolumn}
\usepackage{bm}
\usepackage{umlaute}

\begin{document}

\preprint{ \today}

\title{
Collisional properties of cold spin-polarized metastable neon atoms
}

\author{P.~Spoden}
\affiliation{Institut f\"ur Quantenoptik, Universit\"at Hannover, Welfengarten 1, D-30167 Hannover}
\author{M.~Zinner}
\affiliation{Institut f\"ur Quantenoptik, Universit\"at Hannover, Welfengarten 1, D-30167 Hannover}
\author{N.~Herschbach}
\affiliation{Institut f\"ur Quantenoptik, Universit\"at Hannover, Welfengarten 1, D-30167 Hannover}
\author{W.~J.~van Drunen}
\affiliation{Institut f\"ur Quantenoptik, Universit\"at Hannover, Welfengarten 1, D-30167 Hannover}
\author{W.~Ertmer}
\affiliation{Institut f\"ur Quantenoptik, Universit\"at Hannover, Welfengarten 1, D-30167 Hannover}
\author{G.~Birkl}
\affiliation{Institut f\"ur Quantenoptik, Universit\"at Hannover, Welfengarten 1, D-30167 Hannover}
\affiliation{Institut f\"ur Angewandte Physik, Technische Universit\"at Darmstadt, Schlossgartenstr. 7, D-64289 Darmstadt}

\date{\today}

\begin{abstract}
We measure the rates of elastic and inelastic two-body collisions
of cold spin-polarized neon atoms in the metastable $^3\mathrm{P}_2$ state
for $^{20}$Ne and $^{22}$Ne
in a magnetic trap.
From particle loss, we determine the loss parameter of inelastic 
collisions 
$\beta=6.5(18) \times 10^{-12}\,\mathrm{cm}^3\mathrm{s}^{-1}$ for $^{20}$Ne
and 
$\beta=1.2(3) \times 10^{-11}\,\mathrm{cm}^3\mathrm{s}^{-1}$ for $^{22}$Ne.
These losses 
 are caused by ionizing (i.e. Penning) collisions 
and occur less frequently than for unpolarized atoms.
This proves the suppression of Penning ionization due to spin-polarization.
From cross-dimensional relaxation measurements,
we obtain elastic scattering lengths of
$a=-180(40) \, a_0$ for $^{20}$Ne
and
$a=+150^{+80}_{-50} \, a_0$ for $^{22}$Ne,
 where $a_0=0.0529\,\mathrm{nm}$.
\end{abstract}

\maketitle


The experimental investigation of the collisional interaction
of cold and ultracold atoms has been a driving force for 
important new physics in recent years. 
The successful
demonstration of Bose-Einstein condensation (BEC) in dilute atomic
gases certainly is one of the most prominent examples of this
development \cite{FirstBEC}.
Other examples are the realization of Fermi-degenerate
systems (see \cite{KetterleNature} for an overview) or the generation 
of molecular BECs \cite{MolecularBEC}.
The collisional properties of ultracold atoms critically determine
the achievability of quantum-degenerate systems, 
their  stability, their coherence properties,
 and the characteristics of collective phenomena 
(see \cite{JulienneNature, KetterleNature} for a review)
and thus deserve detailed investigation.
All along,  
experimental work on cold collisions for an additional atomic species
has initiated a new line of exciting research.
Special interest nowadays arises from the investigation of non-alkali systems, 
such as He$^*$ \cite{Orsay,ENS}, Yb \cite{Yb}, Cr \cite{ChromStreulaenge, ChromFeshbach}, 
with their specific interaction properties.

An important novel class of atomic species in this respect are cold and ultracold rare gas atoms 
in metastable triplet states (RG$^*$ atoms).
Due to their high internal energy, these atoms show 
collisional properties and experimental provisions absent in 
other atomic systems.
 Important results have already been achieved in He$^*$,
such as the realization of BEC
\cite{Orsay,ENS}, 
the implementation of 
highly efficient electronic detection 
of quantum-degenerate atomic samples \cite{Orsay}, its application to
 the determination of the elastic scattering length \cite{Seidelin}, and 
the formation of He$^*_2$ dimers  
in purely long-range
states \cite{Dimer}. With this paper, we extend the experimental investigation of 
cold collisions of metastable atoms to the case of neon.

Compared to helium, neon has a more complex internal structure.
This causes for example the occurrence of more than one 
metastable triplet state and of 
anisotropic electrostatic interaction potentials, which influence
the dynamics of elastic and inelastic collisions \cite{EindhovenElastic, NeonLimit}.
Inelastic interactions of RG$^*$ atoms are 
particularly interesting due to their
specific exoergic character.
With high probability, RG$^*$ atoms exhibit
 Penning ionization reactions \cite{Penning},
where one of the atoms is deexcited to the ground state,
the other atom is ionized, and an electron is released.
Penning ionization
may be suppressed if the colliding atoms
are spin-polarized to a spin-stretched state
since then Penning ionization does not conserve the spin
quantum number.
The extent of
suppression depends critically on
anisotropic contributions to the interaction.
In the case of He$^*$ these contributions are small and
 ionization is suppressed by four  to five orders of magnitude  \cite{Seidelin}.
The  heavier rare gases however have stronger
anisotropic contributions, 
 since the excitation of an 
electron to the metastable state creates a $p^5$-core with
non-zero orbital angular momentum.
Indeed, no suppression of Penning ionization was observed in Xe$^*$ \cite{Xenon}.
For neon, Doery et al. calculate suppression ratios between  $1$ and $10^4$, 
which depend most sensitively on exact interaction potentials \cite{NeonLimit}, 
which were not available at the time. Experimental investigations are needed, but
so far measurements of the ionization rate have only been
reported for unpolarized atoms \cite{EindhovenMOT, DrZinner, DrSpoden}. 


In this paper, we present the experimental determination of the 
rates of elastic and inelastic collision of spin-polarized bosonic 
$^{20}$Ne and $^{22}$Ne atoms 
in the $J{=}m_J{=}+2$ substate of the metastable $^3\mathrm{P}_2$ state.
For both isotopes, we determine the elastic scattering length $a$,
the loss parameter of inelastic collisions $\beta$,
and demonstrate the suppression of Penning ionization
due to spin-polarization.

In our experiment, we capture Ne$^*$ atoms 
 in a magneto-optical trap from a Zeeman-decelerated atomic beam \cite{1473}.
After optical pumping into the $m_J{=}+2$ state,
the atoms are transferred into a magnetic Ioffe-Pritchard trap
with radial gradient of $295 \,\mathrm{G/cm}$ 
and axial curvature of $215\, \mathrm{G/cm}^2$.
As a last step of preparation, we apply one-dimensional Doppler cooling \cite{Doppler},
where the trapped atoms 
are irradiated  by two $\sigma^+$ polarized laser beams
 along the symmetry axis of the magnetic trap.
For this purpose, we operate the trap 
with an offset magnetic field of $25 \,\mathrm{G}$. 
The vibrational frequencies, measured with $^{20}\mathrm{Ne}$, are 
 $\omega_x = 2 \pi \times 80(1)\, \mbox{s}^{-1}$ in axial direction
 and $\omega_r = 2 \pi \times 186(1)\, \mbox{s}^{-1}$ in radial direction.
By this preparation we obtain typical temperatures of
 $T_x=450\,\mu\mathrm{K}$ in axial and $T_r=600\,\mu\mathrm{K}$
in radial direction. A typical atom cloud contains $N = 2 \times 10^8$ atoms,
at a mean density $\overline{n} = 4 \times 10^{10} \,\mathrm{cm}^{-3}$,
the latter being the ensemble average of the number density $n(\vec{r})$:
$\overline{n}=N^{-1}\int \!\! n^2(\vec{r}) \, \mbox{d}^3\!r$.
By a Stern-Gerlach experiment, we could verify that the spin-polarization 
persists during Doppler cooling.
We could not detect any atoms in the $J{=}2$, $m_J{\leq}1$ substates,
from which we deduce a lower bound of $95\%$ for the atom population in the
$m_J{=}2$ substate.

As one-dimensional Doppler cooling in the magnetic trap
gives different temperatures, $T_x < T_r$,
 we obtain an ensemble out of equilibrium. 
In each dimension however, potential and kinetic energy are balanced such
that $\omega_{x(r)} \, \sigma_{x(r)}= \sqrt{k_\mathrm{B} T_{x(r)}/m}$, where
$\sigma_{x(r)}$ is the spatial width in axial (radial) direction and $\sqrt{k_\mathrm{B} T_{x(r)}/m}$
 the velocity spread, with
Boltzmann's constant $k_{\mathrm{B}}$, and the atomic mass $m$.
After cooling, the trapped ensemble 
trends back to equilibrium, which can be observed by a changing
aspect ratio of the spatial distribution.
In parallel, the number of atoms decreases with time and their
 mean temperature increases.
We observe heating rates which are density dependent and
range from a few $\mu \mathrm{K}/\mathrm{s}$ up to 
$50\, \mu \mathrm{K}/\mathrm{s}$ for increasing densities.

\begin{figure}[t]
\centering
    \includegraphics[width=170pt, keepaspectratio]{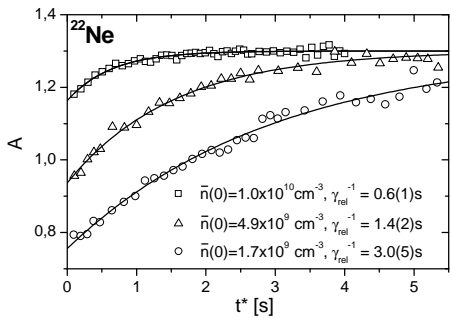} 
\caption{Cross-dimensional relaxation of the aspect ratio of $^{22}$Ne ensembles 
for different initial densities $\overline{n}(0)$.
Higher densities lead to shorter relaxation times $\gamma_{\mathrm{rel}}^{-1}$.
The lines are fits according to Eq. \ref{AspectScaled}.
During relaxation $\overline{n}$ and $\overline{T}$ change.
We compensate for the resulting change in $\gamma_{\mathrm{rel}}$ 
by using the rescaled time   
$t^*(t) =
\int_0^t \left( \overline{n}(t')\, \overline{v}(t')\right) / \left( \overline{n}(0)\, \overline{v}(0) \right) \mbox{d}t'$,
similar to \cite{Oxford}.
}
\label{fig:relaxation}
\end{figure}
In general, collision rates depend on the atom cloud's density and temperature.
We derive $\overline{n}$, $T_x$, and $T_r$
from absorption images 
 which are taken between $0.5$ and $2\, \mathrm{ms}$ after switching off the magnetic trap.
For the determination of these parameters, statistical uncertainties (typically $2-3 \%$) are
 negligible as compared to systematic uncertainties.
Uncertainties are given as one standard deviation and represent the quadrature sums
 of statistical and systematic contributions,
dominated by the systematic uncertainty in density ($ 23\% \leq \Delta \overline{n}/\overline{n} \leq 28\%$).

We determine the rates of elastic collisions from 
cross-dimensional relaxation measurements 
 \cite{Cornell,Oxford,ChromStreulaenge}.
The equilibrating transfer of energy
from the radial to the axial dimension
changes the atom cloud's aspect ratio $A(t)$,
according to
\begin{equation} \label{AspectScaled}
\dot{A}(t)=-\gamma_{\mathrm{rel}}(A(t)-A_{\mathrm{eq}}) \, .
\end{equation}
Here, $A_{\mathrm{eq}}$ is the equilibrium aspect ratio
and $\gamma_{\mathrm{rel}}$ the relaxation rate.
\begin{figure}[t]
\centering
    \includegraphics[width=170pt, keepaspectratio]{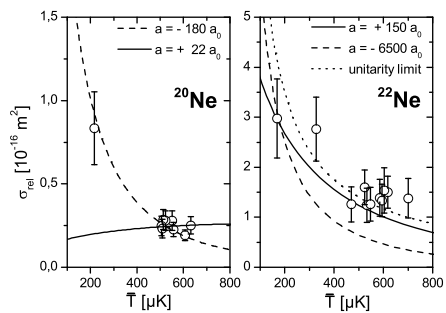} 
\caption{
Effective relaxation cross sections $\sigma_\mathrm{rel}$ 
obtained from cross-dimensional relaxation measurements and calculations:
(left)
For $^{20}$Ne we obtain a scattering length of $a = -180(40) \, \mathrm{a}_0$.
The measurements  between $500$ and $700 \, \mu \mathrm{K}$
are also consistent with a positive $a = +22(5) \, \mathrm{a}_0$;
(right)
The large $\sigma_\mathrm{rel}$-values of  $^{22}$Ne
are close to the unitarity limit.
Our data agree best with theory for 
$a=+150^{+80}_{-50} \, \mathrm{a}_0$. We can rule out negative scattering lengths 
since theory cannot match our data,
as exemplified for $a=-6500 \, \mathrm{a}_0$ (see text).
}
\label{fig:SigmaRel}
\end{figure}
We observe reequilibration of $^{20}$Ne 
and $^{22}$Ne (Fig.~\ref{fig:relaxation})
for different initial densities $\overline{n}(0)$
and find relaxation rates
 which are directly proportional to $\overline{n}(0)$ \cite{DrSpoden}.
 Hence relaxation is predominantly caused by collisions
and ergodic mixing is of no relevance 
for our determination of $\gamma_{\mathrm{rel}}$ \cite{MonteCarlo}.
Therefore, we can describe the relaxation rate of the atomic ensemble
by an effective relaxation cross section $\sigma_\mathrm{rel}$, 
\begin{equation} \label{eq:SigmaRel}
\gamma_\mathrm{rel}=\sigma_\mathrm{rel}\,\overline{n}\, \overline{v}\, ,
\end{equation}
where 
$\overline{v}=\left({16 k_B \overline{T}/\pi m}\right)^{1/2}$
is the average relative velocity of the colliding atoms, with
the mean temperature $\overline{T}=\frac{1}{3}(T_x+2T_r)$.

Since the relaxation process is driven stronger
by collisions with high relative velocity,
Kavoulakis et al.~have derived 
$\sigma_\mathrm{rel}\propto
\langle \sigma_\mathrm{el}(v) \, v \rangle_\mathrm{nst.}$
from a non-standard thermal average of the elastic collision cross section 
$\sigma_\mathrm{el} (v)$,
(Eqs.~(72) and (88) in \cite{Smith}),
where  $v$ denotes the relative velocity of the colliding atoms.
Thereby $\sigma_\mathrm{rel}$
becomes a function of scattering length $a$ and temperature: 
$\sigma_\mathrm{rel}=\sigma_\mathrm{rel}(a,\overline{T})$.

Figure~\ref{fig:SigmaRel} shows the relaxation
cross sections of $^{20}$Ne and $^{22}$Ne as a function of temperature \cite{TrapConfiguration}. 
We have prepared atom clouds at different temperatures either by
varying the efficiency of Doppler cooling, 
or by an additional adiabatic expansion within the trap. 
The temperature assigned to each measurement is the average
of temperatures $\overline{T}(t)$ weighted with
$\dot{A}(t)$. This procedure reflects
 that the period of fastest change in $A(t)$ is most important
for the extraction of $\gamma_\mathrm{rel}$ \cite{AspectHeating}.

In our measurements, relaxation is dominated by s-wave collisions
between spin-polarized atoms, as 
the centrifugal barrier
for d-waves is $k_\mathrm{B} \times 5.6 \, \mathrm{mK}$ in energy. 
In the temperature range we explore,
 the elastic scattering cross section depends
 on both the magnitude $\it{and}$ the sign of the scattering length $a$,
as opposed to the case
of ultracold collisions, where $\sigma_\mathrm{el} \to 8 \pi a^2$.
To analyze the data, we use 
the temperature dependence of $\sigma_\mathrm{rel}(a,\overline{T})$
which we derive from numerical calculations:
we start by calculating $\sigma_\mathrm{el}(v)$ as a function of $v$ for different 
input values of $a$
by solving the radial Schr\"odinger equation. 
With the non-standard average $\langle \sigma_\mathrm{el}(v) \, v \rangle_\mathrm{nst.}$
we calculate $\sigma_\mathrm{rel}(a,\overline{T})$
for temperatures between $100$ and $800\, \mu \mathrm{K}$ (lines in Fig.~\ref{fig:SigmaRel})
and determine the scattering length  
by fitting these curves to the experimental data ($\chi^2$ minimization procedure).

These calculations are based on the
short-range ab-initio potentials 
 given by Kotochigova et al.~\cite{Kotochigova},
and the long-range van-der-Waals potentials given by Derevianko et al.\ \cite{Derevianko}.
We treat the collision process by a single-channel calculation,
instead of including the five potentials involved, for three reasons:
(i) the sensitivity of the $v$-dependence of $\sigma_\mathrm{el}(v)$ on the short-range potentials 
is negligible (see also \cite{ChromStreulaenge, EindhovenElastic});
(ii) the $4\%$ uncertainty in the van-der-Waals coefficients for the different potentials involved
is comparable to their differences \cite{Derevianko};
(iii) the variation in the calculated $\sigma_\mathrm{rel}(a,\overline{T})$ 
caused by these differences 
are negligible when compared to experimental uncertainties
in the measurement of $\sigma_\mathrm{rel}$.

For $^{20}$Ne, 
we obtain a negative scattering length of $a= -180(40) \, \mathrm{a}_0$, 
with Bohr's radius $a_0=0.0529\,\mathrm{nm}$. 
Between 
$500$ and $650\, \mu \mathrm{K}$, our data are also consistent
with a positive value of $a= +22(5) \, \mathrm{a}_0$. 
The sign of the scattering length is determined by
the measurements at $\overline{T} \approx 200\, \mu \mathrm{K}$, 
giving a $74$ times higher probability for 
$a= -180 \, \mathrm{a}_0$ than for $a= +22 \, \mathrm{a}_0$.
For $^{22}$Ne, we obtain  
relaxation cross sections $\sigma_\mathrm{rel}$
which are three to five times larger as compared to $^{20}$Ne
(Fig.~\ref{fig:SigmaRel}). We obtain $a= +150^{+80}_{-50} \, \mathrm{a}_0$. 
Due to the proximity of the $\sigma_\mathrm{rel}$-data to the unitarity limit,
the extracted $a$ is less precise than in the case of $^{20}$Ne.
Our calculations show that the relaxation cross sections for 
$a<0$ do not match our data.
We exemplify this by including a plot of $\sigma_\mathrm{rel}$
for $a= -6500 \, \mathrm{a}_0$ in Fig.~\ref{fig:SigmaRel}. All curves
calculated for $-6500 \, \mathrm{a}_0 < a < 0$ lie below this curve,
and thus disagree with our data.
We  therefore conclude that the scattering length of $^{22}$Ne is positive.
%

\begin{figure}[t]
	\includegraphics[width=170pt, keepaspectratio]{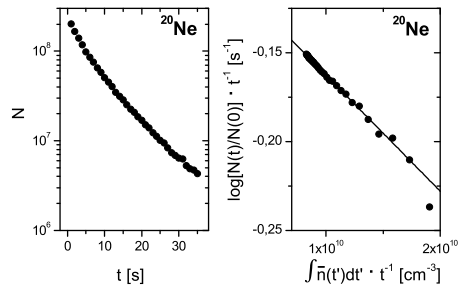}
\caption{
(left) Atom loss of a spin-polarized $^{20}$Ne ensemble.
Initially, $2 \times 10^8$ atoms were trapped at
a mean density $\overline{n}=2 \times 10^{10}\, \mathrm{cm}^{-3}$.
(right) Data presented according to Eq.~(\ref{DecayLinear})
and a linear fit giving
 a one body loss rate $\alpha=(10.3(3) \,\mathrm{s})^{-1}$
and a loss constant $\beta=6.5(18)\times10^{-12}\,\mathrm{cm}^3\mathrm{s}^{-1}$.}
\label{fig:DecayNe20}
\end{figure}

From atom loss in the magnetic trap,
we determine the rate of
binary inelastic collisions of a spin-polarized ensemble.
The decrease in atom number  $N(t)$ (Fig.~\ref{fig:DecayNe20}, left)
stems from one-body losses ($\propto N$) and two-body losses ($\propto N^2$).
We have no signature of higher order loss processes
 in our measurements. 
One body-losses with rate $\alpha$ are mainly caused by the
$14.73\, \mathrm{s}$ lifetime of the $^3\mathrm{P}_2$ state \cite{1473}
 and by background gas collisions.
 Two-body losses 
 depend on the probability of two atoms to collide ($\propto n^2(\vec{r})$)
and on the loss parameter $\beta$. 
The decay of the particle number is described by
\begin{equation} \label{DecayGauss}
\frac{\mbox{d}}{\mbox{d}t}N\!(t)=-\alpha N\!(t) - \beta \frac{N^2(t)}{V_{\mathrm{eff}}(t)} \, ,
\end{equation}
with the effective volume $V_{\mathrm{eff}}=N/\overline{n}$.
We find that $V_{\mathrm{eff}}$ increases due to heating, 
so that an explicit solution of Eq.~(\ref{DecayGauss}) requires
assumptions on the heating mechanisms involved. 
We circumvent this by formally integrating Eq.~(\ref{DecayGauss})
 and obtain a linear relation between quantities,
which are easily derived from measured data:
\begin{equation}  \label{DecayLinear}
\frac{1}{t} \log \frac{N(t)}{N(0)} = - \alpha - \beta \, \left( \frac{1}{t}\, \int_0^t \overline{n}(t')dt' \right) \, .
\end{equation}
The constant offset in
this relation is the one-body loss rate $\alpha$ 
and the slope is the two-body loss parameter $\beta$. 

By presenting the data of a trap loss measurement (Fig. \ref{fig:DecayNe20} (left))
according to Eq.~(\ref{DecayLinear}), we find the expected linear dependence (Fig. \ref{fig:DecayNe20} (right)).
From this, we get the one-body decay rate
 $\alpha^{-1}=10.3(3)\, \mathrm{s}$,
and the two-body loss parameter
 $\beta=6.5(18) \times 10^{-12}\,\mathrm{cm}^3\mathrm{s}^{-1}$.
The uncertainty in $\beta$ is dominated by the systematic uncertainty in $\overline{n}$.
Comparing $\beta$ to the loss parameter of an unpolarized ensemble
 $\beta_{\mathrm{unpol}}=2.5(8) \times 10^{-10}\,\mathrm{cm}^3\mathrm{s}^{-1}$ \cite{DrZinner, DrSpoden},
we obtain the suppression of 
inelastic collisions due to spin-polarization $\beta_{\mathrm{unpol}} / \beta = 38(16)$
for $^{20}$Ne.

We get a second independent measurement of the rate of inelastic collisions
 from observed heating rates. Besides other possible heating mechanisms,
 two-body loss
leads to intrinsic heating: these collisions happen most likely in the center of the trap.
Therefore, the average energy carried away by a lost atom
is less than the mean energy per trapped atom. 
By fitting the resulting heating rate $\dot{\overline{T}}/{\overline{T}}=\beta \overline{n}/4$ \cite{Dalibard2}
to the rising mean temperature $\overline{T}$, we
 obtain $\beta_{\mathrm{heat}} = 7(2) \times 10^{-12}\,\mathrm{cm}^3\mathrm{s}^{-1}$.
This value agrees well with the result from the trap loss measurement 
and indicates that other heating mechanisms do not contribute significantly
to observed heating rates. 

We performed similar measurements with $^{22}$Ne at an initial atom number of $9 \times 10^7$
($\overline{n}=1.4\times 10^{10}\,\mathrm{cm}^{-3}$)
yielding a one-body decay rate of $\alpha^{-1}=14.5(2) \,\mathrm{s}$.
We find a loss parameter of $\beta=1.2(3) \times 10^{-11}\,\mathrm{cm}^3\mathrm{s}^{-1}$
from the decay measurement, and $\beta_{\mathrm{heat}} =1.3(3) \times 10^{-11}\,\mathrm{cm}^3\mathrm{s}^{-1}$
from observed heating.
With a loss parameter $\beta_{\mathrm{unpol}}=8(5)\times 10^{-11}\,\mathrm{cm}^3\mathrm{s}^{-1}$
\cite{DrSpoden} for unpolarized atoms, the suppression of two-body losses
is $\beta_{\mathrm{unpol}} / \beta = 7(5)$ for $^{22}$Ne.

\begin{table}[t]
\caption{\label{tab:Summary}Summary of collision parameters: 
effective relaxation cross section $\sigma_\mathrm{rel}$ for $\overline{T}=200$ and $550 \,\mu \mathrm{K}$,
 scattering length $a$, two-body loss parameters $\beta$ and $\beta_{\mathrm{unpol}}$ for polarized and unpolarized atoms
and their suppression ratio.
}
\begin{ruledtabular}
\begin{tabular}{lrrr}
{ }&{ }& $^{20}$Ne & $^{22}$Ne \\
\hline
\\[-3mm]
$\sigma_\mathrm{rel}(\approx 200 \mu \mathrm{K})$& $[10^{-17}\,\mathrm{m}^2]$ & $8(2)$ &$30(8)$\\
$\sigma_\mathrm{rel}(\approx 550 \mu \mathrm{K})$& $[10^{-17}\,\mathrm{m}^2]$ & $2.8(7)$ &$13(3)$\\
$a$&$[a_0]$		& $-180(40)$		&$+150^{+80}_{-50}$ \\
$\beta$&$[10^{-12}\,\mathrm{cm}^3\mathrm{s}^{-1}]$		&$6.5(18)$&	$12(3)$\\
$\beta_{\mathrm{unpol}}$&$[10^{-12}\,\mathrm{cm}^3\mathrm{s}^{-1}]$ &$250(80) $&	$80(50)$\\
$\beta_{\mathrm{unpol}}/\beta$	& { }	&		$38(16)$  &$7(5)$\\
\end{tabular}
\end{ruledtabular}
\end{table}
In an additional experiment, we simultaneously detected atoms with absorption imaging
and measured
the rate of ions escaping from a cloud of trapped $^{22}$Ne atoms
 with a multichannel plate (MCP) detector.
From these measurements (analogous to \cite{1473})
 we can establish the ratio of the rate of ionizing
inelastic collisions ($\beta_\mathrm{ion}\, \overline{n}$) 
to the total rate of inelastic collisions ($\beta \, \overline{n}$):
$\beta_\mathrm{ion}/\beta = 1.1(2)$, where the uncertainty is dominated by the
calibration uncertainty of the MCP detector.
These measurements confirm, that the observed inelastic collisions 
are caused by Penning ionization to a very high degree ($\geq 90 \%$).

Our measurements prove the suppression of Penning ionization
due to spin-polarization. Compared to He$^*$
this suppression is much less pronounced.
The observed values for $\beta_{\mathrm{unpol}}/\beta$
do not confirm that large suppression ratios are likely \cite{NeonLimit, Prospects}.
We expect that refined calculations of the ionization rate with precise interaction potentials \cite{Kotochigova, Derevianko}
will reproduce the experimental results. 

As a summary, Table~\ref{tab:Summary} gives the collisional parameters 
for both neon isotopes as presented in this paper.
With these measurements, a significant contribution towards the 
understanding of the collisional
properties of spin-polarized metastable neon atoms is gained.
 These results will initiate further theoretical and experimental investigations,
such as detailed calculations of the molecular potentials and their
manipulation by external electromagnetic fields, e.g.\@ to modify
the rate of elastic or inelastic collisions.
Concerning the quest for BEC, the determined ratio
of elastic to inelastic collisions suggests  $^{22}$Ne
to be better suited for evaporative cooling than  $^{20}$Ne.
Accordingly, we have carried out
first experiments of evaporative cooling of $^{22}$Ne
and an increase in
phase space density was already observed \cite{DrSpoden}. 
Whether a BEC of $^{22}$Ne can be realized hereby,
will be revealed by forthcoming evaporation experiments.

We thank A.~Bunkowski and M.~Johanning
 for valuable discussions
and are grateful for financial support by the DFG ({\it Schwerpunktprogramm SPP 1116})
 and the European Commission ({\it RTN Network Cold Quantum Gases}).


\begin{thebibliography}{abc}
\bibitem{TrapConfiguration}
Measurements were taken at magnetic offset fields of $5$, $11$, $15$, and $25 \, \mathrm{G}$.
We have no indication that the relaxation rate changes with the magnetic field 
in this range.
\bibitem{AspectHeating}
Intrinsic heating does not affect relaxation measurements
since $\dot{\sigma}_{x}/\sigma_{x}=\dot{\sigma}_{r}/\sigma_{r}$, which leaves $A$ unchanged.
\end{thebibliography}
\end{document}